\documentclass[sigconf, screen]{acmart}
\usepackage{xcolor} 
\usepackage{multirow}
\usepackage{xspace}
\usepackage{fancybox}
\usepackage{balance}
\usepackage{tikz}
\usetikzlibrary{positioning}

\AtBeginDocument{%
  }

\begin{document}

\balance

\title{Software Testing Beyond Closed Worlds:\\ Open-World Games as an Extreme Case}




\author{Yusaku Kato}
\affiliation{%
  \institution{Ritsumeikan University}
  \city{Ibaraki}
  \country{Japan}}
\email{is0610ks@ed.ritsumei.ac.jp}

\author{Norihiro Yoshida}
\affiliation{%
  \institution{Ritsumeikan University}
  \city{Ibaraki}
  \country{Japan}}
\email{norihiro@fc.ritsumei.ac.jp}

\author{Erina Makihara}
\affiliation{%
  \institution{Ritsumeikan University}
  \city{Ibaraki}
  \country{Japan}}
\email{makihara@fc.ritsumei.ac.jp}

\author{Katsuro Inoue}
\affiliation{%
  \institution{Ritsumeikan University}
  \city{Ibaraki}
  \country{Japan}}
\email{inoue-k@fc.ritsumei.ac.jp}




\begin{abstract}
Software testing research has traditionally relied on closed-world assumptions,
such as finite state spaces, reproducible executions, and stable test oracles.
However, many modern software systems operate under uncertainty, non-determinism,
and evolving conditions, challenging these assumptions.
This paper uses open-world games as an extreme case to examine the limitations
of closed-world testing.
Through a set of observations grounded in prior work, we identify recurring
characteristics that complicate testing in such systems, including inexhaustible
behavior spaces, non-deterministic execution outcomes, elusive behavioral
boundaries, and unstable test oracles.
Based on these observations, we articulate a vision of software testing beyond
closed-world assumptions, in which testing supports the characterization and
interpretation of system behavior under uncertainty.
We further discuss research directions for automated test generation, evaluation
metrics, and empirical study design.
Although open-world games serve as the motivating domain, the challenges and
directions discussed in this paper extend to a broader class of software systems
operating in dynamic and uncertain environments.
\end{abstract}

\ccsdesc[500]{Software and its engineering~Software testing and debugging}
\ccsdesc[300]{Software and its engineering~Search-based software engineering}
\ccsdesc[100]{Computing methodologies~Machine learning}
\keywords{Software testing, Open-world games, Fuzzing}



\acmYear{2026}\copyrightyear{2026}
\setcopyright{cc}
\setcctype[4.0]{by}
\acmConference[FSE Companion '26]{34th ACM Joint European Software Engineering Conference and Symposium on the Foundations of Software Engineering}{July 5--9, 2026}{Montreal, QC, Canada}
\acmBooktitle{34th ACM Joint European Software Engineering Conference and Symposium on the Foundations of Software Engineering (FSE Companion '26), July 5--9, 2026, Montreal, QC, Canada}
\acmDOI{10.1145/3803437.3805587}
\acmISBN{979-8-4007-2636-1/26/07}

\maketitle

\section{Introduction}
\begin{figure*}
  \centering
  \includegraphics[width=\textwidth]{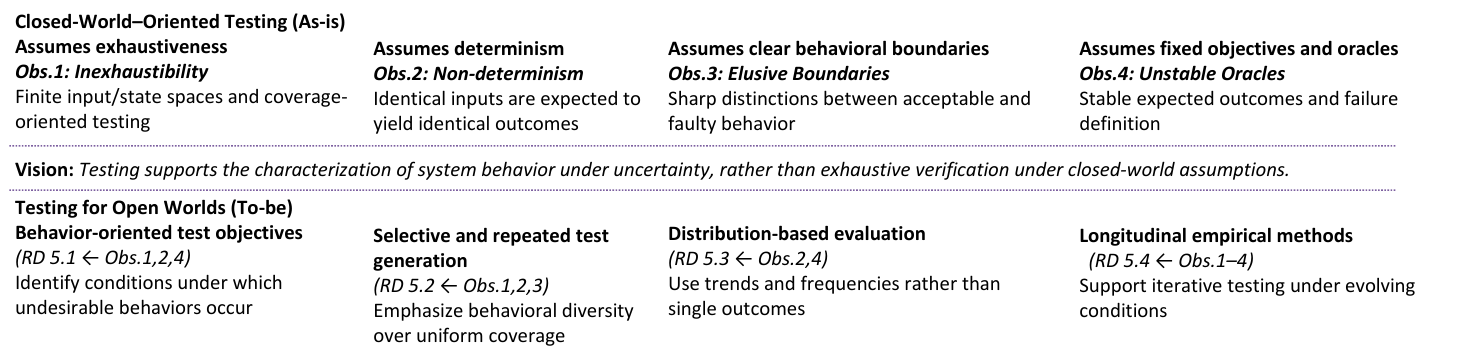}
  \caption{Overview of the paper:
  limitations of closed-world testing assumptions revealed by open-world games (top),
  and corresponding research directions for testing under uncertainty (bottom).}
  \label{fig:vision}
\end{figure*}

Software testing research has traditionally been grounded in closed-world assumptions, under which the system under test is expected to have a finite and stable state space and test executions are assumed to produce reproducible outcomes~\cite{ammann2016introduction,myers2004art,Bertolino2007}.
Under these assumptions, well-established notions such as test coverage, test adequacy, and fault detection effectiveness have enabled systematic evaluation and comparison of testing techniques~\cite{Goodenough1975,Frankl1988}.

However, an increasing number of software systems no longer conform to these assumptions~\cite{Bertolino2007}.
Modern systems often interact with complex environments, incorporate autonomous components, and exhibit behaviors that depend on long execution histories~\cite{whittaker2012google,Calinescu2018}.
Consequently, uncertainty has become an inherent property rather than an exceptional condition~\cite{delemos2013software}.
These characteristics raise fundamental questions about how software testing should be conceptualized and evaluated when reproducibility and stable execution conditions can no longer be assumed~\cite{Fraser2013,Bertolino2007}.

In this paper, we focus on \emph{open-world games} as a representative and extreme
case of such systems.
Open-world games are commonly regarded as games that allow players to freely
explore large virtual environments.
Players can pursue objectives without being constrained by predefined sequences
or linear progressions, emphasizing player-driven exploration and non-linear
interaction~\cite{Alexander2017}.
Such games are typically characterized by vast virtual environments, long-running
executions, and rich interactions among players, environments, and autonomous
entities~\cite{Wuji,Inspector}.
These properties lead to extremely large and dynamic behavior spaces, variable
execution outcomes, and continuously evolving testing targets~\cite{kato2025,Yu2023,Politowski2022}.
Although automated testing techniques have been applied to games and interactive
systems, open-world games make the limitations of testing assumptions derived
from closed-world games particularly explicit~\cite{kato2025,Wuji,Inspector}.

Rather than proposing a new testing technique or tool, this paper takes a step back to examine what open-world games reveal about the nature of software testing under uncertainty.
By treating open-world games as an extreme case, we aim to distill a set of recurring characteristics that challenge conventional testing perspectives.
These characteristics are presented as a set of observations (see Section~\ref{sec:observations}), which highlight structural properties of systems operating beyond closed-world assumptions.
Figure~\ref{fig:vision} provides a high-level overview of this progression,
from observations to research directions.

Based on these observations, we articulate a vision for software testing beyond closed-world assumptions.
Specifically, we argue for a shift away from viewing testing primarily as a means of exhaustive verification or definitive correctness assessment, toward viewing it as an activity that supports the characterization and interpretation of system behavior under varying and uncertain conditions.
We further outline a set of research directions suggested by this perspective for automated test objectives, test generation, evaluation metrics, and empirical study design.

Although open-world games serve as the motivating domain of this paper, the challenges discussed are not limited to games.
Systems such as autonomous driving software, metaverse platforms, and other interactive systems that continuously operate in open and dynamic environments also exhibit uncertainty, non-determinism, and behavioral variability~\cite{Calinescu2018,delemos2013software}.
From this perspective, open-world games can be seen as an extreme case that exposes fundamental limitations of closed-world testing assumptions and motivates a broader rethinking of software testing practices.

\section{Background}
\label{sec:background}

\subsection{Complexity of Open-World Games}
Open-world games allow players to freely explore large virtual environments and to choose actions without being constrained by predefined sequences or linear scenarios~\cite{kato2025, Barbero2025}. In contrast to closed-world games, which typically rely on tightly constrained, level-based or mission-based designs (e.g., platform games such as \emph{Super Mario Bros.} or role-playing games with fixed progression structures such as early entries in the \emph{Final Fantasy} series), open-world games emphasize player-driven exploration and non-linear interaction. 

Unlike classical games with manageable state variables, modern open-world games developed with game engines involve a vast number of variables, leading to state explosion~\cite{kato2025}. System behavior in such games emerges from complex interactions among player actions, environmental dynamics, and multiple in-game entities~\cite{Alexander2017, Prasetya2020}. This emergent nature challenges the testing assumptions that have traditionally been effective for closed and constrained game designs~\cite{Murphy-Hill2014}.

\subsection{Open-World Games as an Extreme Case}
Automated testing, game balance, and player experience evaluations for games have been widely studied~\cite{Politowski2021,Politowski2022,Wuji,Inspector}, with recent studies emphasizing the impact of diverse player behaviors and play styles on exploration and test coverage in complex game environments~\cite{MIMIC_YIFEI_ASE_2025,kato2025}.

Software testing research has also recognized challenges in systems operating under uncertainty, such as autonomous driving and self-adaptive software~\cite{Barr2015, delemos2013software, Dutta2020}. Simulators in autonomous driving often utilize distribution-based evaluations to handle environment variability~\cite{Deng2023}. However, while autonomous driving systems possess rigid traffic rules, open-world games often lack universal oracles. In games, uncertainty is not just a factor to be mitigated but an inherent property of play and emergent interaction. 

Although prior studies have demonstrated the feasibility of applying automated testing techniques to large-scale and non-deterministic game environments~\cite{kato2025,Wuji,Inspector}, they primarily adopt a tool-oriented perspective, focusing on specific techniques or systems. In contrast, this paper treats open-world games as an \emph{extreme case}, a frontier of uncertainty that exceeds the current assumptions in both classical game testing and adjacent domains such as autonomous driving, to derive general observations beyond closed-world assumptions.

\section{Observations} \label{sec:observations}
Figure~\ref{fig:vision} situates the observations discussed in this section
(Obs1--Obs4) within the overall structure of the paper.
The top part of the figure summarizes the limitations of closed-world testing
assumptions revealed by open-world games, which are elaborated through the
following observations.

This section summarizes a set of recurring characteristics observed in the testing of open-world games, as supported by prior studies on automated game testing and interactive systems.

\paragraph{\textbf{Observation 1: Inexhaustibility.}}
In the testing of open-world games, it is observed that behavior spaces are vast and history-dependent.
Execution paths diverge substantially due to differences in player strategies and play styles.
Because of this structural complexity, the space of possible behaviors becomes extremely large.
As a result, exhaustive exploration through testing is infeasible in practice.
This observation is supported by prior studies on automated testing of games and interactive systems.
Fuzzing- and search-based approaches have repeatedly reported diminishing returns when attempting to scale exploration in large and open-ended environments~\cite{kato2025,Politowski2021,Politowski2022}.
Research on player modeling further shows that diverse play styles lead to substantially different execution traces even within the same game context~\cite{bostan2020game,MIMIC_YIFEI_ASE_2025}.
These results collectively reinforce the impracticality of exhaustive testing in open-world settings.

\paragraph{\textbf{Observation 2: Non-determinism.}}
In the testing of open-world games, it is observed that identical inputs or
action sequences can lead to different behaviors across executions.
This variability arises from multiple sources, including autonomous agent
decisions, physics simulations, stochastic events, and concurrent interactions
among in-game entities.
As a result, individual test outcomes are not fully repeatable, even under
nominally identical conditions.
Such non-determinism complicates the interpretation of testing results based
on single executions or binary pass/fail judgments.
Related challenges have been widely documented in prior work on adaptive and
interactive systems, where non-determinism is treated as an intrinsic property
rather than an anomaly~\cite{delemos2013software,Calinescu2018}.
In the context of automated game testing, several studies report substantial
run-to-run variability caused by AI decision-making and physics simulation,
even under controlled experimental settings~\cite{Politowski2021,Prasetya2020,Wuji,Inspector,kato2025}.
These findings support the view that non-determinism is a fundamental
characteristic that testing must explicitly account for in such systems.

\paragraph{\textbf{Observation 3: Elusive Boundaries.}}
In the testing of open-world games, it is observed that clear boundaries between
qualitatively different system behaviors are difficult to identify.
Because input and state spaces are open and continuous, system behavior often
exhibits high sensitivity to small variations.
Minor changes in actions, timing, or configuration parameters may lead to
disproportionate differences in observed behavior.
As a consequence, the distinction between acceptable and problematic behavior
becomes blurred, complicating reasoning about correctness.
This observation is supported by prior research on testing systems with large
or continuous input domains.
Search-based and automated testing studies report that small input variations
can trigger abrupt behavioral changes, making boundary identification difficult~\cite{Fraser2013,Xiao2023}.
Research on game testing and player experience further shows that subtle
behavioral differences can significantly affect perceived quality and user
satisfaction~\cite{bostan2020game,Politowski2022}.
Together, these studies corroborate the elusiveness of behavioral boundaries
in open and interactive systems.
\paragraph{\textbf{Observation 4: Unstable Oracles.}}
In the testing of open-world games, it is observed that expected outcomes and
correctness criteria are inherently difficult to define with precision.
System behavior emerges from complex interactions among players, environments,
and autonomous entities, making exact expected results hard to specify.
Moreover, these expectations tend to change over time due to updates,
balance adjustments, and evolving player practices.
Consequently, test oracles cannot be assumed to be fixed, precise, or stable
throughout the testing process.
This observation is supported by prior research on testing non-deterministic
and non-testable systems.
Patel and Hierons argue that when precise test oracles cannot be defined,
traditional testing approaches break down, motivating alternative mechanisms
such as metamorphic testing, N-version testing, and statistical hypothesis
testing~\cite{Patel2018}.
Extensive work on the oracle problem further shows that correctness judgments
are often approximate and context-dependent, particularly in evolving and
interactive systems~\cite{Barr2015,delemos2013software}.
Related challenges have also been studied in the context of flaky tests,
where test outcomes vary across executions due to inherent sources of
non-determinism.
Dutta et al.\ report that algorithmic non-determinism is a major cause of
flaky tests in machine-learning-based systems, and that developers often cope
by relaxing assertion thresholds or adopting more flexible judgment
criteria~\cite{Dutta2020}.
Research on automated game testing similarly reports oracle instability as
gameplay mechanics and quality expectations change over time~\cite{Politowski2021,kato2025}.
Together, these studies support the difficulty of relying on static and
unambiguous test oracles in such settings.

\section{Vision: Testing Beyond Closed Worlds}
\label{sec:vision}

Taken together, the observations motivate a shift in software testing from exhaustive verification toward the characterization and interpretation of system behavior under uncertainty. 
In this vision, testing is no longer a one-time activity aimed at eliminating uncertainty to confirm correctness. 
Instead, it is a foundational and ongoing process through which practitioners collect empirical evidence about how a system behaves across repeated executions.

This empirical evidence supports the characterization of which behaviors and failures may arise, under what conditions, and with what likelihood. 
By moving away from binary correctness, testing provides the necessary insights to guide decisions on where to focus further testing, debugging, and investigation under limited resources. 
Open-world games serve as a useful lens for this vision because they manifest multiple, structurally-linked testing challenges simultaneously, demanding a fundamental rethinking of software testing beyond closed-world assumptions.

\section{Research Directions}
\label{sec:directions}

The vision presented in Section~\ref{sec:vision} calls for revisiting how automated software testing
research is conducted for systems operating under uncertainty.
Rather than treating uncertainty as a factor to be eliminated,
this perspective views testing as a means to provide empirical evidence
that supports reasoning about system behavior across executions.
In this section, we outline several research directions that follow from this view
and focus on how testing can better support decisions about test generation,
evaluation, and empirical study design under uncertainty.

\subsection{Test Objectives}

Under closed-world assumptions, test objectives are commonly framed in terms of exhaustive coverage
or binary correctness judgments.
In systems characterized by uncertainty, such objectives become less informative,
as identical inputs or scenarios may lead to diverse outcomes and behaviors across executions.

A key research direction is to rethink test objectives so that they emphasize behavioral diversity,
failure modes, and the conditions under which undesirable behaviors are likely to occur.
This shift moves the role of testing away from establishing definitive correctness
and toward supporting decisions about how to prioritize further testing,
debugging, and investigation under uncertainty.

\subsection{Automated Test Generation}

Revisiting test objectives has direct implications for automated test generation.
When exhaustive exploration is infeasible, test generation techniques must
selectively explore behaviorally meaningful regions of the system rather than
attempting uniform coverage.

\paragraph{\textbf{Selective exploration under cost constraints.}}
Test generation strategies should balance execution cost with behavioral diversity.
This may be achieved by leveraging heuristics derived from user interaction patterns,
system objectives, or observed execution histories to prioritize behaviorally
meaningful regions of the input and state spaces.

\paragraph{\textbf{Explicit treatment of execution variability.}}
Repeated and long-running executions should be treated as first-class concerns.
Variability observed across executions should be regarded as valuable evidence for
identifying conditions under which undesirable behaviors are likely to occur,
rather than as noise to be eliminated.

\paragraph{\textbf{Generality across automated testing paradigms.}}
These research directions apply broadly across fuzzing, search-based testing, and
learning-based approaches.
Rather than prescribing a specific technique, they aim to support decisions about
how to allocate testing effort effectively under uncertainty.

\subsection{Evaluation and Metrics}

The proposed vision motivates a reconsideration of how automated testing techniques should be evaluated under uncertainty. Traditional evaluation metrics, such as code coverage or failure counts, implicitly
assume stable and repeatable execution outcomes.

However, in open-world games, evaluation is further complicated by characteristics of the underlying game engines. Engine-level mechanisms such as physics simulation, event scheduling, and real-time update loops introduce inherent variability across executions, even under similar test conditions. As a result, test outcomes cannot always be interpreted using deterministic oracles, and single execution results may provide limited insight into overall system behavior.

Rather than relying solely on scalar metrics, evaluation should support reasoning about behavioral diversity, failure tendencies, and the conditions under which undesirable behaviors are likely to occur. This view is consistent with prior work on the oracle problem, which emphasizes that correctness judgments are often approximate and context-dependent in complex and evolving systems~\cite{Barr2015}. While distribution-based evaluation is utilized in adjacent domains, such as autonomous driving, to assess reliability~\cite{Deng2023}, its application to games poses unique challenges. In the autonomous driving of an existing study, evaluation centers on the violation rates of predefined safety rules~\cite{Deng2023}. In contrast, in games, distributions must characterize the richness of player experience or the stability of emergent system states, as these factors lack universal oracles and are often better captured by metrics of state diversity or transition coverage than by traditional code coverage alone~\cite{IJON}.

This shift also calls for a reinterpretation of reproducibility. Instead of expecting identical outcomes across executions, reproducibility may be understood as the ability to observe consistent patterns or trends that can inform subsequent testing and analysis activities. Such an interpretation better aligns evaluation practices with systems that operate under non-determinism and uncertainty.

To make distribution-based evaluation more conclusive for real-world development decisions, a promising direction is to adopt probabilistic oracles and risk-based thresholding. In this approach, developers define acceptable bounds for behavioral distributions. A test result is considered conclusive if the observed behavior exposes a failure trend that exceeds a predefined threshold based on the frequency of occurrence and its impact on the system. This approach enables developers to prioritize debugging efforts on high-risk anomalies while accepting minor variability as an inherent property of open-world games.

\subsection{Empirical Studies and Benchmarks}

These research directions extend naturally to the design of empirical studies and benchmarks.
Many existing studies rely on fixed programs, fixed inputs, and limited execution runs,
which may be insufficient for systems whose behavior evolves over time and across conditions.

Future empirical research should place greater emphasis on longitudinal observation
and repeated experimentation to capture how behaviors and failure modes manifest over time.
Benchmarks may need to support varying conditions, extended execution periods,
and the systematic collection of behavioral data,
thereby enabling empirical studies that better support decisions about testing strategies
for systems operating under uncertainty.

\section{Conclusion}

This paper used open-world games as an extreme case to examine the limitations of
closed-world testing assumptions.
Through a set of observations, we highlighted structural characteristics such as
inexhaustible behavior spaces, non-deterministic execution outcomes, elusive
behavioral boundaries, and unstable test oracles, which collectively challenge
traditional notions of exhaustive and deterministic testing.

Motivated by these observations, we argued for a reconceptualization of software
testing that shifts the focus from exhaustive verification to supporting the
characterization and interpretation of system behavior under uncertainty.
We outlined research directions for test objectives, automated test generation,
evaluation metrics, and empirical study design that better align testing practice
with such conditions.

Although open-world games served as the motivating domain, the challenges discussed
are not limited to games.
As software systems increasingly operate in open, dynamic, and interactive
environments, the perspective presented in this paper points toward broader
methodological shifts in automated software testing research.

\begin{acks}
This work was supported by JSPS KAKENHI Grant Number JP24K02923. We thank Shinnosuke Iwatsubo and Hiroki Mukai at Ritsumeikan University for their comments.

\end{acks}

\newpage
\bibliographystyle{ACM-Reference-Format}
\bibliography{reference}

@ARTICLE{Barr2015,
  author={Barr, Earl T. and Harman, Mark and McMinn, Phil and Shahbaz, Muzammil and Yoo, Shin},
  journal={IEEE Transactions on Software Engineering}, 
  title={The Oracle Problem in Software Testing: A Survey}, 
  year={2015},
  volume={41},
  number={5},
  pages={507-525},
  doi={10.1109/TSE.2014.2372785}}

@INPROCEEDINGS{kato2025,
  author={Kato, Yusaku and Yoshida, Norihiro and Makihara, Erina and Inoue, Katsuro},
  booktitle={2025 32nd Asia-Pacific Software Engineering Conference (APSEC)}, 
  title={BiFuzz: A Two-Stage Fuzzing Tool for Open-World Video Games}, 
  year={2025},
  volume={},
  number={},
  pages={1017-1020},
  doi={10.1109/APSEC66846.2025.00120}
}

@book{ammann2016introduction,
  author = {Ammann, P. and Offutt, J.},
  description = {Introduction to Software Testing - Paul Ammann, Jeff Offutt - Google Books},
  isbn = {9781107172012},
  lccn = {2016032808},
  publisher = {Cambridge University Press},
  series = {Introduction to Software Testing},
  title = {Introduction to Software Testing},
  url = {https://books.google.de/books?id=bQtQDQAAQBAJ},
  year = 2016
}

@ARTICLE{Goodenough1975,
  author={Goodenough, John B. and Gerhart, Susan L.},
  journal={IEEE Transactions on Software Engineering}, 
  title={Toward a theory of test data selection}, 
  year={1975},
  volume={SE-1},
  number={2},
  pages={156-173},
  doi={10.1109/TSE.1975.6312836}}

@ARTICLE{Frankl1988,
  author={Frankl, P.G. and Weyuker, E.J.},
  journal={IEEE Transactions on Software Engineering}, 
  title={An applicable family of data flow testing criteria}, 
  year={1988},
  volume={14},
  number={10},
  pages={1483-1498},
  doi={10.1109/32.6194}}

@book{myers2004art,
  title={The art of software testing},
  author={Myers, Glenford J and Badgett, Tom and Thomas, Todd M and Sandler, Corey},
  volume={2},
  year={2004},
  publisher={Wiley Online Library}
}

@INPROCEEDINGS{Bertolino2007,
  author={Bertolino, Antonia},
  booktitle={Future of Software Engineering (FOSE '07)}, 
  title={Software Testing Research: Achievements, Challenges, Dreams}, 
  year={2007},
  volume={},
  number={},
  pages={85-103},
  doi={10.1109/FOSE.2007.25}}

@book{whittaker2012google,
  title={How Google tests software},
  author={Whittaker, James A and Arbon, Jason and Carollo, Jeff},
  year={2012},
  publisher={Addison-Wesley}
}

@ARTICLE{Calinescu2018,
  author={Calinescu, Radu and Weyns, Danny and Gerasimou, Simos and Iftikhar, Muhammad Usman and Habli, Ibrahim and Kelly, Tim},
  journal={IEEE Transactions on Software Engineering}, 
  title={Engineering Trustworthy Self-Adaptive Software with Dynamic Assurance Cases}, 
  year={2018},
  volume={44},
  number={11},
  pages={1039-1069},
  doi={10.1109/TSE.2017.2738640}}

@inproceedings{delemos2013software,
author="de Lemos, Rog{\'e}rio
and Giese, Holger
and M{\"u}ller, Hausi A.
and Shaw, Mary
and Andersson, Jesper
and Litoiu, Marin
and Schmerl, Bradley
and Tamura, Gabriel
and Villegas, Norha M.
and Vogel, Thomas
and Weyns, Danny
and Baresi, Luciano
and Becker, Basil
and Bencomo, Nelly
and Brun, Yuriy
and Cukic, Bojan
and Desmarais, Ron
and Dustdar, Schahram
and Engels, Gregor
and Geihs, Kurt
and G{\"o}schka, Karl M.
and Gorla, Alessandra
and Grassi, Vincenzo
and Inverardi, Paola
and Karsai, Gabor
and Kramer, Jeff
and Lopes, Ant{\'o}nia
and Magee, Jeff
and Malek, Sam
and Mankovskii, Serge
and Mirandola, Raffaela
and Mylopoulos, John
and Nierstrasz, Oscar
and Pezz{\`e}, Mauro
and Prehofer, Christian
and Sch{\"a}fer, Wilhelm
and Schlichting, Rick
and Smith, Dennis B.
and Sousa, Jo{\~a}o Pedro
and Tahvildari, Ladan
and Wong, Kenny
and Wuttke, Jochen",
editor="de Lemos, Rog{\'e}rio
and Giese, Holger
and M{\"u}ller, Hausi A.
and Shaw, Mary",
title="Software Engineering for Self-Adaptive Systems: A Second Research Roadmap",
bookTitle="Software Engineering for Self-Adaptive Systems II: International Seminar, Dagstuhl Castle, Germany, October 24-29, 2010 Revised Selected and Invited Papers",
year="2013",
publisher="Springer Berlin Heidelberg",
address="Berlin, Heidelberg",
pages="1--32",
isbn="978-3-642-35813-5",
doi="10.1007/978-3-642-35813-5_1",
url="https://doi.org/10.1007/978-3-642-35813-5_1"
}

@ARTICLE{Fraser2013,
  author={Fraser, Gordon and Arcuri, Andrea},
  journal={IEEE Transactions on Software Engineering}, 
  title={Whole Test Suite Generation}, 
  year={2013},
  volume={39},
  number={2},
  pages={276-291},
  doi={10.1109/TSE.2012.14}}

@INPROCEEDINGS{Wuji,
  author={Zheng, Yan and Xie, Xiaofei and Su, Ting and Ma, Lei and Hao, Jianye and Meng, Zhaopeng and Liu, Yang and Shen, Ruimin and Chen, Yingfeng and Fan, Changjie},
  booktitle={2019 34th IEEE/ACM International Conference on Automated Software Engineering (ASE)}, 
  title={Wuji: Automatic Online Combat Game Testing Using Evolutionary Deep Reinforcement Learning}, 
  year={2019},
  volume={},
  number={},
  pages={772-784},
  doi={10.1109/ASE.2019.00077}}

@INPROCEEDINGS{Inspector,
  author={Liu, Guoqing and Cai, Mengzhang and Zhao, Li and Qin, Tao and Brown, Adrian and Bischoff, Jimmy and Liu, Tie-Yan},
  booktitle={2022 IEEE Conference on Games (CoG)}, 
  title={Inspector: Pixel-Based Automated Game Testing via Exploration, Detection, and Investigation}, 
  year={2022},
  volume={},
  number={},
  pages={237-244},
  doi={10.1109/CoG51982.2022.9893630}}

@INPROCEEDINGS{MIMIC_YIFEI_ASE_2025,
  author={Chen, Yifei and Habchi, Sarra and Wei, Lili},
  booktitle={2025 40th IEEE/ACM International Conference on Automated Software Engineering (ASE)}, 
  title={MIMIC: Integrating Diverse Personality Traits for Better Game Testing Using Large Language Model}, 
  year={2025},
  volume={},
  number={},
  pages={39-51},
  doi={10.1109/ASE63991.2025.00012}
}

@INPROCEEDINGS{Yu2023,
  author={Yu, Jiongchi and Wu, Yuechen and Xie, Xiaofei and Le, Wei and Ma, Lei and Chen, Yingfeng and Hu, Jingyu and Zhang, Fan},
  booktitle={2023 IEEE/ACM 45th International Conference on Software Engineering (ICSE)}, 
  title={GameRTS: A Regression Testing Framework for Video Games}, 
  year={2023},
  volume={},
  number={},
  pages={1393-1404},
  doi={10.1109/ICSE48619.2023.00122}}

@INPROCEEDINGS{Politowski2022,
  author={Politowski, Cristiano and Guéhéneuc, Yann-Gaël and Petrillo, Fabio},
  booktitle={2022 IEEE/ACM 6th International Workshop on Games and Software Engineering (GAS)}, 
  title={Towards Automated Video Game Testing: Still a Long Way to Go}, 
  year={2022},
  volume={},
  number={},
  pages={37-43},
  doi={10.1145/3524494.3527627}}

@InProceedings{Barbero2025,
author="Barbero, Giulio
and M{\"u}ller-Brockhausen, Matthias
and Preuss, Mike",
editor="Anutariya, Chutiporn
and Bonsangue, Marcello M.
and Budhiarti-Nababan, Erna
and Sitompul, Opim Salim",
title="Challenges of Open World Games for AI: Insights from Human Gameplay",
booktitle="Data Science and Artificial Intelligence",
year="2025",
publisher="Springer Nature Singapore",
address="Singapore",
pages="127--141",
isbn="978-981-97-9793-6",
doi = {https://doi.org/10.1007/978-981-97-9793-6_9}
}

@inproceedings{Alexander2017,
author = {Alexander, Ryan and Martens, Chris},
title = {Deriving quests from open world mechanics},
year = {2017},
isbn = {9781450353199},
publisher = {Association for Computing Machinery},
address = {New York, NY, USA},
url = {https://doi.org/10.1145/3102071.3102098},
doi = {10.1145/3102071.3102098},
booktitle = {Proceedings of the 12th International Conference on the Foundations of Digital Games},
articleno = {12},
numpages = {7},
location = {Hyannis, Massachusetts},
series = {FDG '17}
}

@inproceedings{Prasetya2020,
author = {Prasetya, I. S. W. B. and Voshol, Maurin and Tanis, Tom and Smits, Adam and Smit, Bram and Mourik, Jacco van and Klunder, Menno and Hoogmoed, Frank and Hinlopen, Stijn and Casteren, August van and Berg, Jesse van de and Prasetya, Naraenda G.W.Y. and Shirzadehhajimahmood, Samira and Ansari, Saba Gholizadeh},
title = {Navigation and exploration in 3D-game automated play testing},
year = {2020},
isbn = {9781450381017},
publisher = {Association for Computing Machinery},
address = {New York, NY, USA},
url = {https://doi.org/10.1145/3412452.3423570},
doi = {10.1145/3412452.3423570},
booktitle = {Proceedings of the 11th ACM SIGSOFT International Workshop on Automating TEST Case Design, Selection, and Evaluation},
pages = {3–9},
numpages = {7},
location = {Virtual, USA},
series = {A-TEST 2020}
}

@inproceedings{Murphy-Hill2014,
author = {Murphy-Hill, Emerson and Zimmermann, Thomas and Nagappan, Nachiappan},
title = {Cowboys, ankle sprains, and keepers of quality: how is video game development different from software development?},
year = {2014},
isbn = {9781450327565},
publisher = {Association for Computing Machinery},
address = {New York, NY, USA},
url = {https://doi.org/10.1145/2568225.2568226},
doi = {10.1145/2568225.2568226},
booktitle = {Proceedings of the 36th International Conference on Software Engineering},
pages = {1–11},
numpages = {11},
location = {Hyderabad, India},
series = {ICSE 2014}
}

@INPROCEEDINGS{Politowski2021,
  author={Politowski, Cristiano and Petrillo, Fabio and Guéhéneuc, Yann-Gaël},
  booktitle={2021 IEEE/ACM International Conference on Automation of Software Test (AST)}, 
  title={A Survey of Video Game Testing}, 
  year={2021},
  volume={},
  number={},
  pages={90-99},
  doi={10.1109/AST52587.2021.00018}}

@book{bostan2020game,
  title={Game user experience and player-centered design},
  author={Bostan, Barbaros},
  year={2020},
  publisher={Springer},
  doi = {https://doi.org/10.1007/978-3-030-37643-7}
}

@inproceedings{Xiao2023,
author = {Xiao, Dongwei and Liu, Zhibo and Wang, Shuai},
title = {PhyFu: Fuzzing Modern Physics Simulation Engines},
year = {2024},
isbn = {9798350329964},
publisher = {IEEE Press},
url = {https://doi.org/10.1109/ASE56229.2023.00054},
doi = {10.1109/ASE56229.2023.00054},
booktitle = {Proceedings of the 38th IEEE/ACM International Conference on Automated Software Engineering},
pages = {1579–1591},
numpages = {13},
location = {Echternach, Luxembourg},
series = {ASE '23}
}

@article{Patel2018,
author = {Patel, Krishna and Hierons, Robert M.},
title = {A mapping study on testing non-testable systems},
year = {2018},
issue_date = {December  2018},
publisher = {Kluwer Academic Publishers},
address = {USA},
volume = {26},
number = {4},
issn = {0963-9314},
url = {https://doi.org/10.1007/s11219-017-9392-4},
doi = {10.1007/s11219-017-9392-4},
journal = {Software Quality Journal},
month = dec,
pages = {1373–1413},
numpages = {41}
}

@inproceedings{Dutta2020,
author = {Dutta, Saikat and Shi, August and Choudhary, Rutvik and Zhang, Zhekun and Jain, Aryaman and Misailovic, Sasa},
title = {Detecting flaky tests in probabilistic and machine learning applications},
year = {2020},
isbn = {9781450380089},
publisher = {Association for Computing Machinery},
address = {New York, NY, USA},
url = {https://doi.org/10.1145/3395363.3397366},
doi = {10.1145/3395363.3397366},
booktitle = {Proceedings of the 29th ACM SIGSOFT International Symposium on Software Testing and Analysis},
pages = {211–224},
numpages = {14},
location = {Virtual Event, USA},
series = {ISSTA 2020}
}

@INPROCEEDINGS{IJON,
  author={Aschermann, Cornelius and Schumilo, Sergej and Abbasi, Ali and Holz, Thorsten},
  booktitle={2020 IEEE Symposium on Security and Privacy (SP)}, 
  title={Ijon: Exploring Deep State Spaces via Fuzzing}, 
  year={2020},
  volume={},
  number={},
  pages={1597-1612},
  doi={10.1109/SP40000.2020.00117}}

@ARTICLE{Deng2023,
  author={Deng, Yao and Zheng, Xi and Zhang, Tianyi and Liu, Huai and Lou, Guannan and Kim, Miryung and Chen, Tsong Yueh},
  journal={IEEE Transactions on Software Engineering}, 
  title={A Declarative Metamorphic Testing Framework for Autonomous Driving}, 
  year={2023},
  volume={49},
  number={4},
  pages={1964-1982},
  doi={10.1109/TSE.2022.3206427}}

\end{document}